\documentclass[sigconf]{acmart}

\AtBeginDocument{%
  }

\setcopyright{acmlicensed}
\copyrightyear{2025}
\acmYear{2025}
\acmDOI{XXXXXXX.XXXXXXX}

\acmConference[EDBT '25]{Barcelona, Spain}
\acmISBN{978-1-4503-XXXX-X/18/06}

\begin{document}

\title{[Experiments \& Analysis] Hash-Based vs. Sort-Based Group-By-Aggregate: A Focused Empirical Study [Extended Version]}

\author{Gaurav Vaghasiya}
\affiliation{
  \institution{Santa Clara University}
  \city{Santa Clara}
  \state{CA}
  \country{USA}
}
\email{gvaghasiya@scu.edu}

\author{Shiva Jahangiri}
\affiliation{
  \institution{Santa Clara University}
  \city{Santa Clara}
  \state{CA}
  \country{USA}
}
\email{sjahangiri@scu.edu}

\begin{abstract}

Group-by-aggregate (GBA) queries are integral to data analysis, allowing users to group data by specific attributes and apply aggregate functions such as sum, average, and count. Database Management Systems (DBMSs) typically execute GBA queries using either sort- or hash-based methods, each with unique advantages and trade-offs. Sort-based approaches are efficient for large datasets but become computationally expensive due to record comparisons, especially in cases with a small number of groups. In contrast, hash-based approaches offer faster performance in general but require significant memory and can suffer from hash collisions when handling large numbers of groups or uneven data distributions. This paper presents a focused empirical study comparing these two approaches, analyzing their strengths and weaknesses across varying data sizes, datasets, and group counts using Apache AsterixDB. Our findings indicate that sort-based methods excel in scenarios with large datasets or when subsequent operations benefit from sorted data, whereas hash-based methods are advantageous for smaller datasets or scenarios with fewer groupings. Our results provide insights into the scenarios where each method excels, offering practical guidance for optimizing GBA query performance.
  
\end{abstract}




\keywords{Group-By, Hash, Sort, Aggregation, AsterixDB}


\maketitle
\section{Introduction}
Group-by and aggregate operations are fundamental to data analysis, allowing for the categorization of data by specific attributes and the application of aggregate functions like sum, average, and count. These operators enable users to identify trends and patterns in large datasets and are crucial in DBMSs, supporting key applications such as reporting, business intelligence, and decision support. Given the widespread use of Group-By queries across industries and their high computational cost, optimizing their execution is critical, especially as datasets grow in size and complexity. 

Traditional methods for executing GBA queries typically fall into two categories: sort-based and hash-based approaches. Sort-based Group-By methods involve sorting data by grouping attributes and then scanning the sorted data to apply aggregate functions. While effective for large datasets, this approach becomes computationally expensive as dataset size increases but the number of groups does not, particularly when grouping by fields like strings with common prefixes. The sorting step often becomes a bottleneck, requiring extensive CPU and I/O resources, leading to longer execution times.

Hash-based Group-By methods, on the other hand, use a hash table to partition data based on grouping attributes. This approach can be more efficient in execution time, especially when input data fits inside the memory. However, hash-based methods pose challenges, including significant memory overhead for maintaining hash tables and intermediate aggregates, potential performance issues from hash collisions, and reliance on data distribution and hashing function design, which can lead to unpredictable performance across different scenarios. It has been seen that certain scenarios favor one method over the other, emphasizing the necessity of supporting both hash- and sort-based algorithms in query-processing systems.

In this paper, we conduct a comprehensive experimental evaluation of both sort-based and hash-based approaches for executing GBA queries. Our study focuses on identifying the conditions under which each approach performs optimally on modern hardware by testing them against varying data sizes, numbers of groups, data types, and distributions. Using Apache AsterixDB as the evaluation platform, we examine the strengths and weaknesses of each method, with a particular focus on execution time, memory usage, and disk I/O behavior. The results provide practical insights for optimizing GBA query execution in large-scale database systems.

The remainder of this paper is organized as follows. Section \ref{background} provides background on GBA operations and outlines the fundamentals of sort-based and hash-based approaches, along with a discussion of Apache AsterixDB’s architecture and its support for these methods. Section \ref{design} describes the design and implementation details of sort-based and hash-based Group-By operations, specifically focusing on how they are executed in AsterixDB. Section \ref{experiments} presents a comprehensive set of experiments to evaluate both approaches. Section \ref{relatedWork} reviews related work in the optimization of Group-By queries, highlighting the key contributions and techniques in prior studies. Section \ref{conclusion} concludes the paper by summarizing our findings and suggesting future research directions to further enhance the efficiency and scalability of GBA queries.

\section{Background}\label{background}
\subsection{Sort and Hash-based GBA}

The sort-based approach involves sorting the data by the group fields before applying the aggregation functions. This method clusters rows with the same group field values together, allowing the aggregation functions to be applied in a single pass over the sorted data. The operation is quite similar to the external sort operation, with the primary difference being that the Group-By operator generates aggregate field results per group fields' values. 

The hash-based approach uses a hash table to track groups (keys) and their corresponding aggregations (values) as data is scanned. Each unique Group-By key is hashed to a slot in the table, where the aggregates are updated. This method bypasses the costly sorting step and is generally faster for datasets with a smaller number of groups \cite{AsterixDBManual,DBLP:journals/tods/DoGN22}. For each incoming record, the hash Group-By operator computes a hash value for the group field and checks if a partial aggregate result for that group already exists. If it does, the record is aggregated into the result; otherwise, a new aggregate entry is created. When the dataset is too large to fit all aggregate records in memory, the system spills the aggregated records to disk, using recursive hashing and aggregation to process and send the data from temporary files.

\subsection{Apache AsterixDB}
Apache AsterixDB \cite{asterixdb,DBLP:journals/pvldb/AlsubaieeAABBBCCCFGGHKLLOOPTVWW14,DBLP:journals/spe/KimBBBBCHJJLLMP20} is an open-source, parallel, shared-nothing big data management system (BDMS) built to support the storage, indexing, modifying, analyzing, and querying of large volumes of semi-structured data.

\begin{figure}[h]
\centerline{\includegraphics[width=0.80\linewidth]{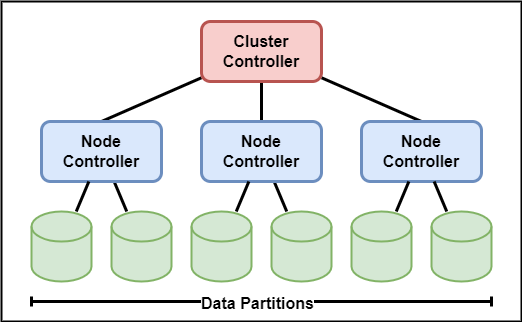}}
\caption{AsterixDB's Architecture}
\label{fig:asterixdbarch}
\vspace{-5mm}
\end{figure}

AsterixDB's entry point for user requests is its Cluster Controller (CC) node, which compiles and transforms the requests into executable jobs. Its Node Controllers (NCs) are the worker nodes that execute the jobs sent by the CC. One NC also serves as the Meta Data Controller Node and provides access to AsterixDB's metadata. Each NC manages one or more data partitions (DPs) as shown in Figure \ref{fig:asterixdbarch}. An instance of each query will be executed in parallel on each data partition that the query needs to access. 

AsterixDB supports optional hints and parameters for users to control how resources, including memory and CPU, will be used. Users may use memory budget parameters to define the maximum possible memory allocation to each instance of a memory-intensive operator (join, sort, etc.) involved in a query. Users may also provide hints to select the physical operator for certain logical operations. AsterixDB's default Group-By operator is sort-based, but the hash-based operator is used if the corresponding hint is provided \cite{AsterixDBManual}. 

AsterixDB provides strong support for modeling and querying both open and closed datasets, which is ideal for semi-structured data \cite{AsterixDBManual}. In closed datasets, the schema is strictly defined, and all records conform to it. Both open and closed records are stored in binary-encoded AsterixDB Data Model (ADM) format \cite{DBLP:journals/pvldb/AlsubaieeBBHKCDL14}, with declared fields stored in metadata for optimized performance and storage efficiency \cite{DBLP:conf/bigdataconf/PirzadehCW15}. In contrast, open datasets do not require a predefined schema. Records can include additional, undeclared fields, offering flexibility for varying attributes within the same dataset. Declared fields are stored in metadata, while the information about undeclared fields of each record is stored within the record itself. This allows applications to introduce new attributes without restructuring the schema.

We chose Apache AsterixDB as our platform for evaluating traditional GBA approach techniques for several reasons. Primarily, our goal was to perform an in-depth, controlled analysis of the two most prevalent Group-By-Aggregate algorithms—hash-based and sort-based methods—under a variety of conditions. AsterixDB's open-source nature allowed us to delve into the internal workings of these algorithms, which is often not feasible with other systems. Additionally, AsterixDB is a parallel big data management system designed for managing and processing large volumes of structured and semi-structured data. AsterixDB uses SQL++ as its declarative query language designed for properly querying semi-structured and nested data. AsterixDB's Data Model (ADM) is a superset of JSON, enabling records of a dataset to be structured or flexible as the user sees fit similar to Couchbase Server and MongoDB. As a parallel BDMS, AsterixDB follows the shared-nothing architecture and supports multi-core and parallel query processing in each cluster node which makes it comparable with other parallel DBMSs such as Postgres. Other features such as stage-based query execution and memory management, support of various SQL operations and their parallel disk-based execution, and Volcano-style execution make AsterixDB's design to be comparable with current modern DBMS and hence make our results and findings extendable to other systems. Features such as stage-based query execution, operator algorithms, and the iterator model are common across systems like PostgreSQL and SQL Server. By carefully designing our experiments to minimize the influence of AsterixDB-specific configurations (e.g., number of NCs, number of data partitions in each NC, operator-specific memory budget allocations), we focused on the fundamental behaviors of the hash-based and sort-based approaches. Consequently, we believe that our findings are not only applicable to AsterixDB but also offer valuable insights relevant to other similar database systems employing similar GBA algorithms.

\section{Designs of GBA Algorithms}\label{design}
\subsection{Sort-based GBA}

\begin{figure}[h]
\centerline{\includegraphics[width=\linewidth]{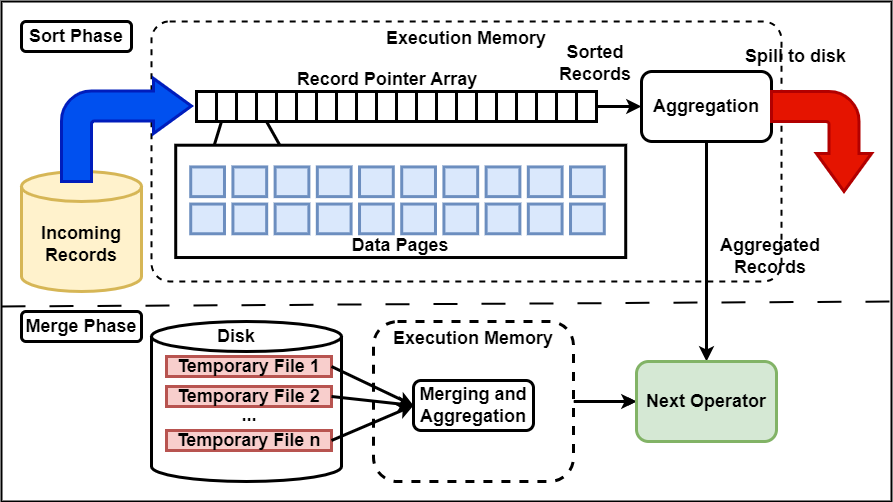}}
\caption{Workflow of Sort-Based Group-By for GBA queries}
\label{fig:sort-group-by}
\vspace{-4mm}
\end{figure}

In AsterixDB, sort-based grouping serves as a fundamental mechanism to efficiently handle GBA queries. AsterixDB uses an external merge sort algorithm to sort the data based on the specified group fields. The operation of sort-based grouping in Apache AsterixDB is structured into two phases as shown in Figure \ref{fig:sort-group-by}, Sort and Merge. 
During the sort phase, records are loaded into memory page by page until the memory is full or all data has been read, then sorted based on group field values. AsterixDB uses normalized keys for sorting, as comparing normalized binary representations is faster than comparing actual values. It also employs a record pointer array, containing the normalized key and the memory location of records, which eliminates the need to move full records in memory \cite{DBLP:journals/csur/Graefe06}. However, when using open types, the actual records are sorted rather than the normalized representation.

After sorting, the records are sent to the aggregation operation, which processes each record sequentially by updating aggregate fields like sums and counts. Since the records are sorted, managing aggregation states is simplified by tracking only the current and previous group values. If memory is sufficient, the aggregated records are passed to the next operator; otherwise, AsterixDB writes them to disk to free up memory and start processing other records. This process continues until all records in the data partition are processed.

The Merge phase consolidates fragmented aggregates from multiple sorted run files into a final output. The Group-By operator assigns an input frame to each temporary file, reads records, and maintains aggregate values based on group field values. If there are more files than input frames, merging occurs recursively in multiple rounds. The final result is either returned as the query output or used for further processing. This sort-based approach is both simple and effective, handling large datasets efficiently.

\begin{figure}[h]
\centerline{\includegraphics[width=\linewidth]{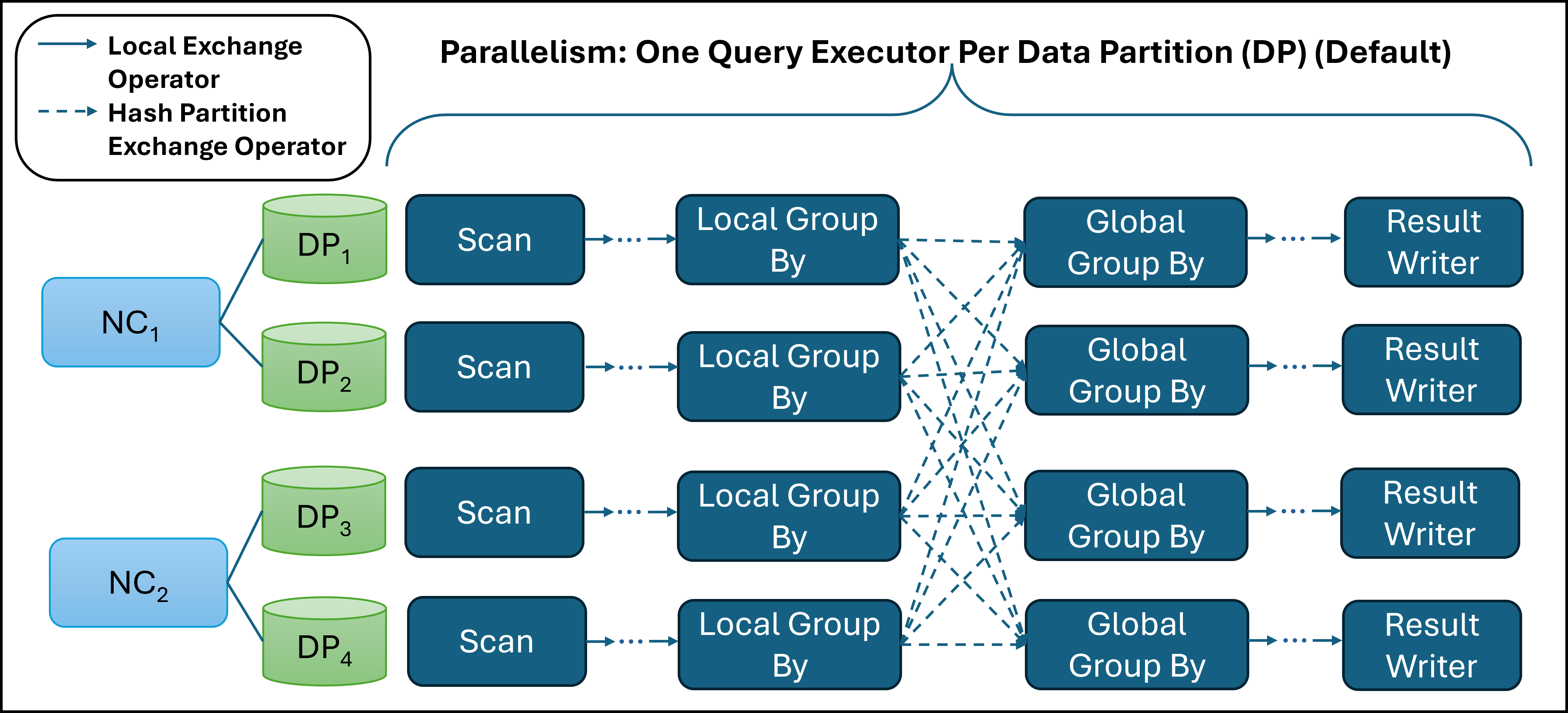}}
\caption{Execution plan for GBA in AsterixDB}
\label{fig:query-plan-HPE}

\end{figure}

In AsterixDB, the execution plan for GBA queries involves two Group-By operators with an implicit aggregation step, separated by a hash-partition exchange operator, as shown in Figure \ref{fig:query-plan-HPE}. The hash-partition exchange distributes data across partitions based on hash values computed from grouping attributes in the GBA case, balancing the load across resources. At each NC, it hashes the data and transfers it to the appropriate partition, enabling efficient parallel processing. This plan applies to both sort- and hash-based GBA methods, with the main difference being the implementation of the Group-By operators. The process begins with ``local aggregation," where each partition aggregates results from its available data.

After local aggregation, the Hash-Partition Exchange operator redistributes records across nodes and partitions using Group-By fields as hash keys, ensuring identical fields are placed together. The redistributed records are then processed by a second Group-By operator for final parallel aggregation on each NC. We refer to this phase as ``global aggregation". The results are then sent to the CC and delivered to the user.

\subsection{Hash-based GBA}

In AsterixDB, hash-based Group-By uses a hash function to process each incoming record for execution of GBA queries. This function determines the bucket or partition where each record's data will be aggregated based on specific fields which in our case are ``group fields". The hash function creates a hash value for group fields, and records are aggregated accordingly in their respective partitions.

The process is divided into two phases to efficiently manage memory and handle large datasets. In the first phase, the operator calculates the number of partitions based on the input data and available memory\cite{DBLP:journals/tods/Shapiro86} to minimize spilling. Each record's hash value is computed, and if an aggregate already exists in memory, the new data is added to it; otherwise, a new aggregate is created. This phase enables fast in-memory aggregation, minimizing disk access. If the aggregated data exceeds the memory capacity, the operator spills some partitions to the disk to free up space to continue processing.

\begin{figure}[h]
\centerline{\includegraphics[width=0.90\linewidth]{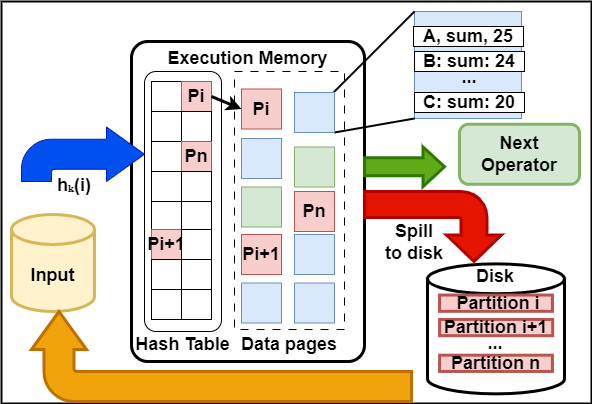}}
\caption{Workflow of Hash-Based Group-By for GBA queries}
\label{fig:hash-group-by}
\vspace{-4mm}
\end{figure}

After processing all records, phase 2 begins. The operator sends the in-memory partitions to the next operator as it will not need any spill partitions to generate final results. Then it goes to phase 1 again and processes all spilled partitions sequentially as shown in Figure \ref{fig:hash-group-by}.

The hash table uses a data structure design that includes a data partition table and a hash table. The data partition table stores the actual records being aggregated, while the hash table holds pointers to these records, facilitating quick location and retrieval of data across data pages. This separation enhances CPU cache efficiency, as the smaller hash table size allows for frequent access, and also at the time of spilling hash table will stay in memory only may spill to the disk. A detailed explanation of the architecture and design of the hash tables of AsterixDB can be found in \cite{DBLP:journals/spe/KimBBBBCHJJLLMP20} and \cite{DBLP:phd/basesearch/Kim18a}.

\section{Experiments}\label{experiments}

This section presents an analysis of experiments evaluating the effectiveness of sort-based and hash-based approaches for GBA queries in Apache AsterixDB. The experiments assess performance across dimensions such as data size, aggregation types, and data characteristics. We describe the configurations and datasets used, followed by the results and insights into the strengths and weaknesses of each approach.

\subsection{Datasets and Configurations}
For a comprehensive and in-depth analysis, we utilized three different benchmarks. TPC-H\cite{tpch} and TPC-DS\cite{tpcds} , widely recognized as the industry standards for evaluating analytical queries, were used to represent realistic workloads. In addition to these, we employed a modified version of the Wisconsin Benchmark\cite{DBLP:conf/sigmod/Jahangiri21} along with its JSON Data Generator\cite{wisconsin-gen}. This allowed us to explore the behavior of GBA queries in a more controlled environment, facilitating a focused analysis of the conditions that significantly impact the performance of different algorithms. While TPC-H and TPC-DS offer realistic scenarios, they often involve complex queries with multiple operators, making it difficult to isolate the performance of individual operators like GBA. The Wisconsin Benchmark, modified to include fields for defining the number of groups, strings of varying sizes, and distributions, allowed us to tailor experiments to specific conditions for deeper analysis.

The experiments are conducted on an AsterixDB cluster setup that includes a CC node and two NC nodes, with each NC managing two data partitions. Also, each NC runs on an Intel NUC machine equipped with 8-core Intel i7 processors and 64 GB of RAM. Each machine utilizes an SSD for storage, ensuring fast data access having sequential write and random write speeds of 4400 MB and 2400 MB per second. Also, all nodes of the cluster are connected via a 2.5 Gbps Ethernet network with a peer-to-peer connection latency under 1 ms. For each NC we limit the JVM memory of AsterixDB to 8GB per node and out of that memory given to GBA operations will be 64 MB per data partition.

We executed each query 11 times, and the average execution time of the last 10 iterations was recorded to exclude the effects of cold-cache starts in the first iteration. To further reduce the impact of data caching, a full scan query was run after each experiment on a dataset five times larger than the available memory. While we run a full scan after each experiment to reduce cache effects, the first execution of a query can still be impacted by a cold cache start at various system levels, including operating system and hardware caches, which are not entirely cleared by the scan. Our goal was to mimic a realistic scenario where the DBMS serves concurrent users, and therefore, some portions of the data may already reside in the cache while others may not. Including the first execution—which represents an extreme case where none of the data is cached—could skew the average execution time and make it unrepresentative of typical performance. By discarding the first run, averaging the subsequent executions, and limiting memory, we aim to obtain a more accurate measure of steady-state performance that reflects the common operating conditions of the system. Additionally, external programs were used to block excess memory on each node, limiting the available memory to approximately 16 GB for AsterixDB and the operating system. This memory limit ensured that queries could run without memory overflow issues, while still providing enough resources for accurate performance measurement.

\subsection{Expr. 1: TPC-H and TPC-DS Benchmarks}
In this experiment, we evaluate the performance of both the sort-based and hash-based GBA approaches using the TPC-H and TPC-DS benchmarks to simulate realistic datasets and query workloads. For both benchmarks, we use a scale factor of 100 to generate the data. Data is loaded to the cluster consisting of one CC and two NC nodes, with each NC managing two data partitions.

\begin{figure}[h]
\centerline{\includegraphics[width=1.1\linewidth]{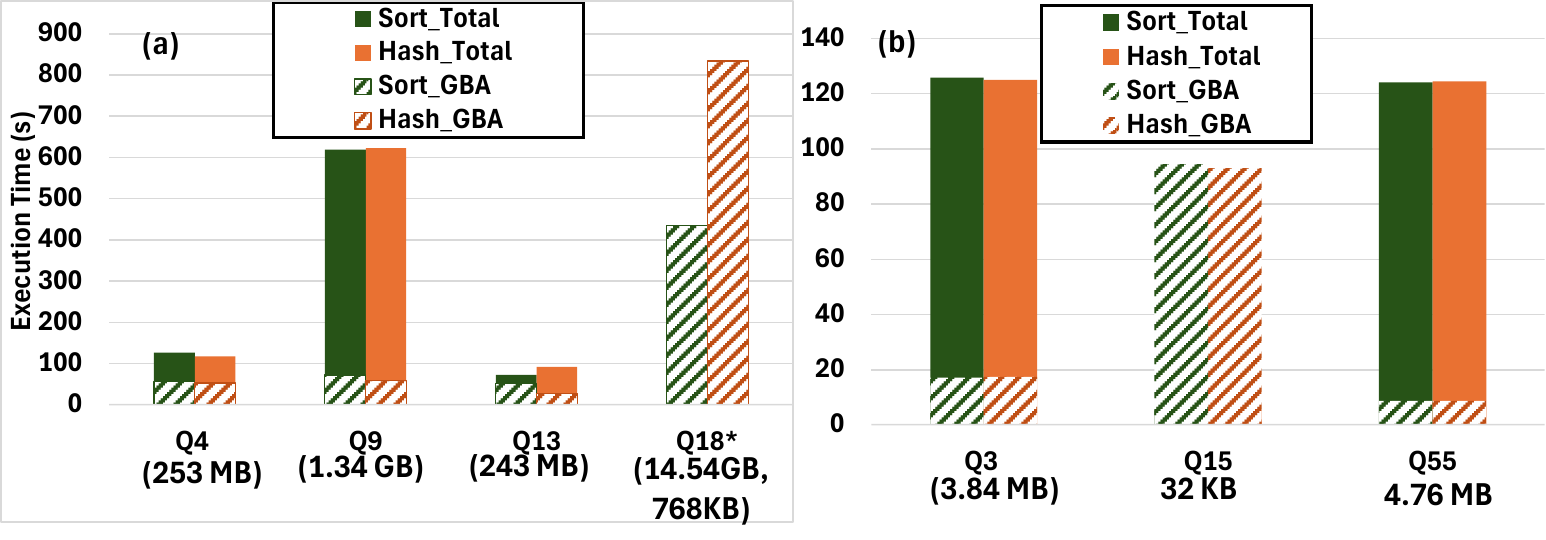}}
\caption{Expr. 1 - (a) TPC-H, (b) TPC-DS. *Q18 has 2 Group-By operators.}
\label{fig:Gtpcds}
\vspace{-4mm}
\end{figure}

To evaluate performance for TPC benchmarks, the query parameters were randomly selected from the benchmark's substitution parameters in each iteration of the query execution. In Figure \ref{fig:Gtpcds}-(a) and (b), the average execution time is plotted on the y-axis, with the query IDs shown on the x-axis. We selected TPC-H Queries 4, 9, 13, and 18 alongside TPC-DS Queries 3, 15, and 55 for their simpler designs and minimal use of complex operators, that contain Group-By-Aggregate operations and fewer computational factors like joins and subqueries to better isolate and observe the impact of GBA algorithms on query performance.


As shown in Figures \ref{fig:Gtpcds}-(a) and (b), in the case of all queries, the hash-based approach performs similarly to sort-based GBA or worse in case of a high amount of data going through the GBA.

Notably, the size of data entering the Group-By operator is quite negligible and small for all TPC-DS queries compared to group memory which is 128MB for the whole configuration. TPC-H queries load more data into the Group-By operator, but in the case of queries Q4, Q9, and Q13 still, it is relatively small compared to the size of the base relations. Also, in Q9 and Q13 GBA time for sort is more compared to hash-based but in the case of those queries output GBA is required to be sorted for the next operator because the AsterixDB query plan optimization sort approach gives sorted output to avoid extra operator complexity. As shown in the overall both approach performs equally for those queries. The Q18 query plan translates to two GBA operations in which the first operation is most impactful as data going in GBA is significant which is around 14.5 GB. In that case, sort-based methods have better execution times than hash-based methods. Due to the inherent complexity of the TPC-H and TPC-DS queries, which involve multiple operators such as joins, subqueries, and order-by clauses, as well as the sophisticated optimization techniques employed by DBMSs, isolating and specifically analyzing the performance of the Group-By Aggregate (GBA) operator proved to be challenging. These benchmarks are designed to evaluate overall system performance rather than the performance of individual operators.

Consequently, the impact of the GBA operator is often minimized by the overhead introduced by scanning large datasets and executing memory- and CPU-intensive operations. Additionally, the limited control over data distributions and query complexity within these benchmarks restricts the ability to create focused experiments that highlight the performance characteristics of GBA operators. As a result, it was not feasible to clearly and specifically study the performance of the GBA operator without the confounding effects of other operators. This limitation motivated our transition to the Wisconsin Benchmark's Dataset and Data Generator, which provided the necessary flexibility to design controlled experiments that accurately assess the performance of both hash-based and sort-based GBA approaches.

In the next experiments, we use a modified Wisconsin Benchmark dataset to replicate different conditions for GBA queries, allowing for a more detailed performance analysis of both approaches.
\subsection{Expr. 2: Impact of Aggregation Types}

In this experiment, we evaluate the scalability and performance of both sort-based and hash-based GBA operators for COUNT and SUM aggregation types, with the dataset size progressively increasing from 2 GB to 64 GB. 
\begin{figure}[h]
\centerline{\includegraphics[width=\linewidth]{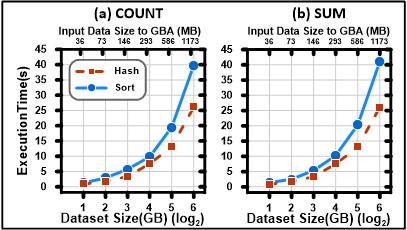}}
\caption{Expr. 2 - Impact of Aggregation Types}
\label{fig:Gten}
\end{figure}

\begin{figure}[h]
\centerline{\includegraphics[width=\linewidth]{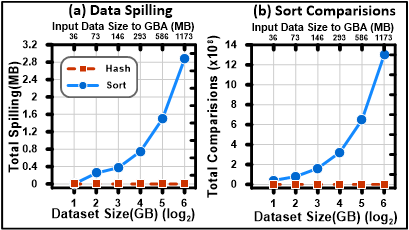}}
\caption{Expr. 2 - Spilling and Sort Comparisons for `COUNT'}
\label{fig:Gten-spill}
\end{figure}
The data is generated using the Wisconsin Data Generator\cite{wisconsin-gen} and loaded into datasets defined with AsterixDB's open data type, with a UUID as the primary key \cite{AsterixDBManual}. The UUID is auto-generated by AsterixDB with random values, ensuring an even distribution of data across partitions. Figure \ref{fig:Gten} presents the experiment results, where the lower x-axis shows the base dataset sizes in gigabytes (logarithmic scale) from 2GB to 64GB, and the upper x-axis indicates input sizes in megabytes for the GBA operator.  The y-axis highlights execution time, with performance trends for both sort-based and hash-based approaches. The structure of the queries for this experiment is as follows:

\begin{verbatim}
    USE wisconsin_dataverse;
    SELECT attribute, COUNT(attribute)
    FROM wisconsin_dataset GROUP BY attribute;
\end{verbatim}

Figure \ref{fig:Gten}-(a) illustrates the GBA operation on integer values from 0 to 9, using COUNT as the aggregation function. This setup yields a limited grouping scenario, producing only 10 output groups. Figure \ref{fig:Gten-spill} represents the amount of data spilled to the disk (MB) and the total number of comparisons during the sort operations for each approach with the `COUNT' aggregation type. The y-axis in Figure \ref{fig:Gten-spill}-a represents the amount of data spilled to the disk (MB) and in Figure \ref{fig:Gten-spill}-b represents the total number of comparisons conducted during the sort operation in the multiple of $10^{8}$. Given the memory configuration of the total of 256 MB of Group-by memory for the whole configuration, the hash-based methods perform entirely in memory leading to near-zero spilling and comparisons compared to sort-based and, achieving better execution times. In contrast, the sort-based approach incurs additional computational overhead from sorting comparisons as shown in Figure \ref{fig:Gten-spill}-(b), resulting in higher execution time. While the amount of data being spilled by the sort-based approach is small due to the small number of groups as shown in Figure \ref{fig:Gten-spill}-(a), the sort comparisons contribute to less efficient performance in this scenario.

Figure \ref{fig:Gten}-(b) shows the result of the same experiment but with SUM as the aggregate function to reveal any difference in execution time that could stem from variance in aggregate type. This comparison helps illustrate that while the type of aggregation function changes, the underlying efficiency of each GBA approach remains consistent.
In summary, the experiment demonstrates that the hash Group-By approach delivers better execution time compared to the sort operator in this setting. With only 10 groups, the hash-based method fits all data in memory, whereas the sort-based approach retains full records, leading to memory exhaustion and data spilling to disk. Furthermore, the choice between count and sum aggregation functions does not significantly impact performance, as both methods rely on the same aggregation state structure \cite{DBLP:journals/spe/KimBBBBCHJJLLMP20}.

\subsection{Expr. 3: Impact of Data Types}
In this series of experiments, we build on the queries from Expr. 2, shifting the focus to Group-By operations involving different data types and variations of strings. This allows us to evaluate how the nature of the data, particularly with varying string characteristics, affects the performance of hash- and sort-based GBA methods.
\subsubsection{\textbf{String vs Integer Comparison}}
\begin{figure}[h]
\centerline{\includegraphics[width=\linewidth]{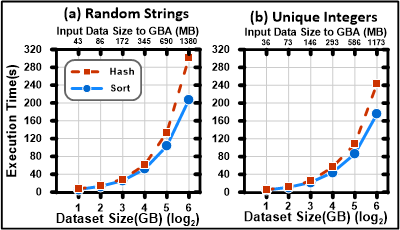}}
\caption{Expr. 3 - Impact of Data Types}
\label{fig:Gexp 2-1}
\end{figure}

\begin{figure}[h]
\centerline{\includegraphics[width=\linewidth]{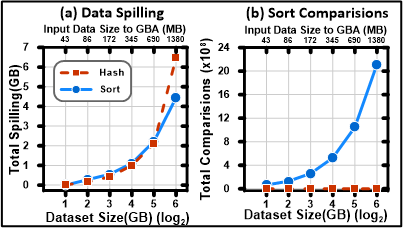}}
\caption{Expr. 3 - Spilling and Sort Comparisons for `Random Strings'}
\label{fig:Gexp 2-spill}
\end{figure}

\begin{figure}[h]
\centerline{\includegraphics[width=\linewidth]{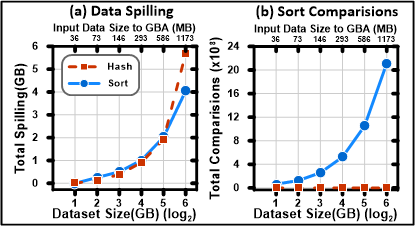}}
\caption{Expr. 3 - Spilling and Sort Comparisons for `Unique Integers'}
\label{fig:Gexp 2-1-spill}
\end{figure}

In this experiment, we explore performance differences for GBA queries across string and integer data types. In Figure \ref{fig:Gexp 2-1}-(a), Group-by is performed on a string attribute with unique and randomly generated values, each 10 characters long. Similarly, in Figure \ref{fig:Gexp 2-1}-(b), Group-by will be performed on randomly generated unique integer values, mimicking the high number of groups typically challenging for Group-By operations as it leads to more comparison and higher amount of data passing through GBA. The aggregation function used in both GBA queries is `COUNT'. Given that the string values are unique, the number of groups generated matches the number of records maintaining comparability with integer data by ensuring a high number of unique group keys. The goal of this experiment is to evaluate how execution time varies under the high number of groups when transitioning from integers to strings. Figures \ref{fig:Gexp 2-spill} and \ref{fig:Gexp 2-1-spill} illustrate the amount of data spilling and the number of comparisons conducted as part of the sorting operations for each approach when handling Group-By fields with unique strings and unique integers respectively, as previously presented in Figure \ref{fig:Gexp 2-1}. In Figures \ref{fig:Gexp 2-spill}-(a) and \ref{fig:Gexp 2-1-spill}-(a), the y-axis represents the amount of data spilled to the disk during the GBA operations while in Figures \ref{fig:Gexp 2-spill}-b and \ref{fig:Gexp 2-1-spill}-b the y-axis shows the number of comparisons conducted by the sort operations. As Figure \ref{fig:Gexp 2-1} shows, the GBA operators having an integer field with unique values as their grouping field have a lower execution time compared to when the grouping field is a string attribute with unique values. This performance difference is due to the higher memory requirements of string values compared to the integer values which can lead to more runs and a slightly higher amount of data spillings to the disk. Specifically, each string value consumes 4 more bytes of memory than integer values during execution of GBA for this case, leading to faster consumption of allocated group memory and more frequent spilling as shown in Figures \ref{fig:Gexp 2-1-spill}-(a) and \ref{fig:Gexp 2-spill}-(a). Additionally,  hash-based GBA spills more data to disk compared to sort-based in both cases due to the hash table overhead taking about \( 1/3^\text{rd} \) of the given memory. This high overhead is due to the high number of groups leading to more hash entries in hash tables from unique group field values.

Next, we study the performance of both GBA methods using variations in string values for the Group-By field.

\subsubsection{\textbf{Exploring various types of strings}}

\begin{figure}[h]
\centerline{\includegraphics[width=\linewidth]{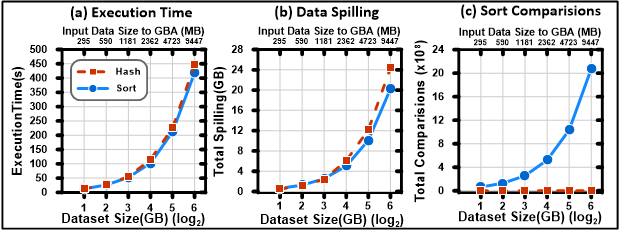}}
\caption{Expr. 3-Impact of Variable Length Strings}
\label{fig:Gvstring}
\end{figure}

The subsequent experiment evaluates GBA algorithms using a string attribute with variable-length values as the grouping field. As illustrated in Figure \ref{fig:Gvstring}, the string lengths range from 10 to 200 characters, with 20\% of the records having lengths between 10 and 50, while the remaining 80\% have lengths between 100 and 200. This experiment aims to evaluate the ability of GBA algorithms to effectively handle variable-length strings, which present significant processing challenges due to their non-uniform distribution and higher storage requirements.

Figures \ref{fig:Gvstring}-(a), -(b), and -(c) present the results of this experiment in terms of execution time, the amount of data spilled by each GBA operator, and the total number of comparisons performed during the sorting operations, respectively. In these figures, the x-axis represents dataset size on a logarithmic scale, ranging from 2GB to 64GB, while the y-axis shows execution time in seconds, data spilling in gigabytes (GB), and the number of sort comparisons, respectively, in Figures\ref{fig:Gvstring}11-(a), -(b), and -(c).

As shown in Figure \ref{fig:Gvstring}-(b), the overall spilling is slightly higher for the hash-based GBA approach compared to the sort-based approach, contributing to its higher execution time. Another factor influencing execution time is the handling of spilled files: the sort-based GBA processes and merges multiple temporary files during its merging phase, whereas the hash-based GBA processes the spilled files sequentially, one at a time. While larger strings impose challenges due to the complexity of storing these records in memory, their variable length does not significantly impact overall performance as in sort-based when records are stored in memory during sorting records compared character by character which does not really add complexity in storing of records in memory. The same for the hash-based approach as the hash table structure and actual records stored in memory during GBA execution are separate it will manage memory efficiently compared to storing records into an in-memory hash table. 

\begin{figure}[h]
\centerline{\includegraphics[width=\linewidth]{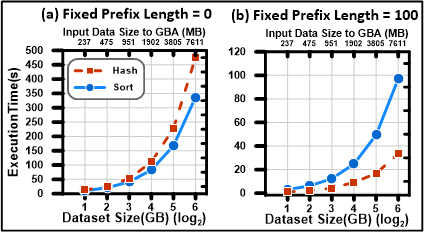}}
\caption{Expr. 3-Impact of various types of Strings}
\label{fig:Gexp 2-2}
\end{figure}

Figure \ref{fig:Gexp 2-2} explores a different scenario where the grouping field values are 100-character strings with two distinct common prefix lengths: 0 characters and 100 characters, as illustrated in Figures \ref{fig:Gexp 2-2}-(a) and -(b). In this case, the input dataset is classified as an Open dataset, with the normalized key feature disabled for the sort-based GBA. This forces character-by-character comparisons during the sorting phase, providing insight into the algorithms’ performance under varying prefix similarity conditions.

This setup aims to assess the computational overhead of sorting in the sort-based GBA method, especially when group keys are only a few—a worst-case scenario for sorting. While this is an engineered condition that may be uncommon in real datasets, it effectively highlights the maximum performance difference between the sort-based and hash-based approaches. As shown in Figure \ref{fig:Gexp 2-2}-(b), sort-based GBA experiences significantly longer execution times due to the increased number of comparisons required for similar strings. In contrast, hash-based GBA efficiently groups identical values using a hash function, with its performance. The hash tables fit into memory for identical strings, and hash-based GBA maintains consistent performance, while the sort-based method becomes increasingly inefficient as prefix string length grows, widening the performance gap.

\subsection{Expr. 4: Impact of Number of Groups}
Next, we analyze the effect of varying the number of group keys on GBA query performance. The goal of this experiment is to evaluate how both approaches handle different group counts, as this can significantly impact GBA performance, particularly in resource-constrained environments. As the number of groups increases, the system’s ability to efficiently manage memory and processing resources is put to the test.

We progressively increase the number of group keys from 10 to $10^{6}$ while keeping the memory allocation for the GBA operator fixed and maintaining a constant dataset size for each experiment. The dataset sizes used are 16 GB, 32 GB, and 64 GB.

\begin{figure}[h]
\centerline{\includegraphics[width=\linewidth]{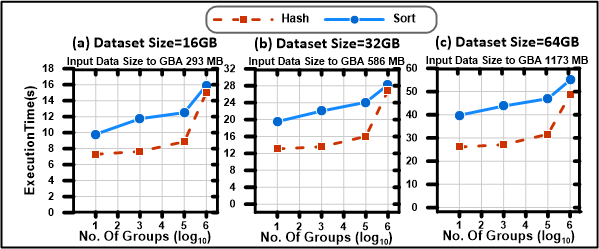}}
\caption{Expr. 4 - Impact of number of groups}
\label{fig:Ggroups}
\end{figure}

Figures \ref{fig:Ggroups}-(a), (b), and (c) present the results, with the x-axis representing the number of groups (logarithmically scaled from 10 to $10^{6}$) and the y-axis showing the average execution time of the GBA queries. The query structure remains consistent with earlier experiments, but the integer grouping field is modified to produce varying numbers of groups.

\begin{figure}[h]
\centerline{\includegraphics[width=\linewidth]{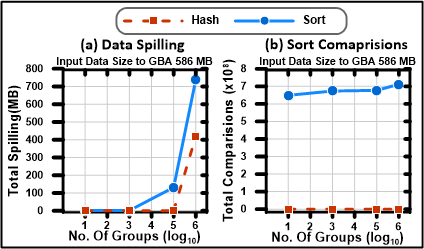}}
\caption{Expr. 4 - Spillings and Sort Comparisons for Various Number of Groups}
\label{fig:Ggroups-spill}
\end{figure}

We anticipated that the sort-based approach would face challenges with sorting and spilling as the number of groups increased, while the hash-based approach was expected to encounter memory limitations, resulting in higher disk I/O.

Figures \ref{fig:Ggroups-spill}-(a) and (b) provide additional insights for this experiment, showing the amount of data spilled to disk and the number of record comparisons performed during the sort operation. The x-axis represents the number of groups, scaled logarithmically from 10 to $10^{6}$, while the y-axis shows data spilling in megabytes (MB) and sort comparisons, respectively, for a dataset size of 32GB.

As illustrated in Figure \ref{fig:Ggroups-spill}-(a), the hash-based approach experiences significant performance degradation as the number of groups increases. This is due to the in-memory hash table becoming overwhelmed, resulting in excessive disk spilling and I/O overhead. These factors lead to additional runs and a marked increase in execution time. When the number of groups reaches $10^{6}$, the hash-based approach encounters a sharp rise in both spilling and execution time, reflecting the limitations of this method under high group cardinality.

In contrast, the sort-based approach exhibits a steady increase in execution time as the group count grows. This behavior is attributed to the escalating sorting and I/O costs associated with larger group sizes, which affect performance in a more predictable and gradual manner.

\section{Related Work}\label{relatedWork}
Sorting and hashing have long been two primary methods widely used by the database community to implement and optimize the performance of various operators. In early relational database systems, only sort-based algorithms were employed. However, over time, hash-based algorithms gained popularity and are now often considered superior for certain use cases. Sorting-based algorithms remained of interest to researchers due to their sorted output, which can benefit subsequent operations requiring sorted input, a concept known as ``interesting order". However, with the introduction of hash-based algorithms, some hash-based methods outperformed sort-based ones. For example, Sort-Merge was outperformed by Hybrid Hash Join \cite{DBLP:conf/icde/Graefe94,DBLP:journals/pvldb/BalkesenATO13, DBLP:journals/vldb/HaasCLS97}.

Sorting and hashing have been extensively used for designing operators such as joins \cite{DBLP:journals/tods/Shapiro86,DBLP:conf/vldb/KitsuregawaNT89, DBLP:conf/vldb/NakayamaK88,DBLP:journals/ife/Graefe12, DBLP:conf/icde/Graefe94}, duplicate removal \cite{DBLP:journals/tods/DoGN22, DBLP:journals/tods/BittonD83}, and grouping and aggregation \cite{DBLP:journals/tods/DoGN22, DBLP:journals/ife/Graefe12}. Their performance has been a subject of study for many researchers \cite{DBLP:journals/pvldb/BalkesenATO13, DBLP:journals/tkde/GraefeLS94,DBLP:conf/sigmod/0002SLLF15}. Traditionally, duplicate removal was performed as a post-step after sorting. However, the authors of \cite{DBLP:conf/vldb/Harder77} demonstrated in System R \cite{DBLP:journals/tods/AstrahanBCEGGKLMMPTWW76} that duplicates can be eliminated during the sorting process, rather than afterward.

There has been much discussion regarding the similarities between sort- and hash-based query processing algorithms. Some research suggests that their costs are similar, differing by small percentages rather than by orders of magnitude \cite{DBLP:journals/tkde/GraefeLS94, DBLP:journals/csur/Graefe93}. Some even argue that hashing is equivalent to sorting by hash value \cite{DBLP:conf/sigmod/0002SLLF15}. However, studies have shown that special cases exist where one method outperforms the other, underscoring the importance of supporting both hash- and sort-based algorithms in query-processing systems \cite{DBLP:journals/tkde/GraefeLS94}. Selecting the optimal Group-By algorithm depends on precise statistics, a task that has proven challenging. As a result, research has shifted towards developing algorithms that are less dependent on statistics, while still maintaining competitive performance. 

In summary, while both sorting and hashing have been extensively studied and optimized, their relative performance in varying conditions, particularly in large-scale parallel DBMS environments such as AsterixDB, warrants further empirical analysis. This paper aims to fill this gap by providing detailed experimental results and practical guidance.

\vspace{-3mm}
\section{Conclusion and Future Work}\label{conclusion}
This paper provides an empirical comparison of hash-based and sort-based GBA query methods using Apache AsterixDB. Our experiments showed that hash-based approaches perform well with smaller group counts and in-memory datasets, avoiding sorting overhead, but struggle with high group cardinality and data skew. In contrast, sort-based methods, while computationally heavier, excel with larger group counts and handle memory spills more efficiently, making them suitable for memory-constrained environments.

The experiments also revealed that data types and complexity impact performance, with sort-based methods struggling with complex strings, while hash-based methods remain unaffected. Overall, hash-based methods have been shown to be preferable for small datasets or low group counts, while sort-based methods perform better in high-cardinality or memory-limited scenarios. These findings provide practical guidance for optimizing GBA query performance based on data characteristics.

Future research could explore adaptive query execution, dynamically switching between hash- and sort-based approaches based on real-time resource usage and data characteristics, optimizing performance without relying on predefined statistics. A possible direction is developing a hybrid approach that combines hashing and sorting within a single query, adapting during runtime. These advancements can help improve the efficiency, scalability, and adaptability of GBA queries, ensuring optimal performance in diverse database environments.

\section*{Acknowledgement}
Special thanks to Michael J. Carey and Wail Y. Alkowaileet for their invaluable feedback and to the Apache AsterixDB team for their continuous support and recommendations.
\clearpage

\bibliographystyle{plain}
\bibliography{sample-sigconf-lualatex}

\begin{thebibliography}{10}

\bibitem{asterixdb}
{Apache AsterixDB}.
\newblock \url{https://asterixdb.apache.org/}.
\newblock 2021.

\bibitem{tpcds}
{TPC-DS Benchmark}.
\newblock \url{https://www.tpc.org/tpcds/}.
\newblock 2021.

\bibitem{tpch}
{TPC-H Benchmark}.
\newblock \url{https://www.tpc.org/tpch/}.
\newblock 2021.

\bibitem{wisconsin-gen}
{Wisconsin Benchmark Data Generator}.
\newblock \url{https://github.com/shivajah/JSON-Wisconsin-Data-Generator}.
\newblock 2021.

\bibitem{DBLP:journals/pvldb/AlsubaieeAABBBCCCFGGHKLLOOPTVWW14}
Sattam Alsubaiee, Yasser Altowim, Hotham Altwaijry, Alexander Behm, Vinayak~R. Borkar, Yingyi Bu, Michael~J. Carey, Inci Cetindil, Madhusudan Cheelangi, Khurram Faraaz, Eugenia Gabrielova, Raman Grover, Zachary Heilbron, Young{-}Seok Kim, Chen Li, Guangqiang Li, Ji~Mahn Ok, Nicola Onose, Pouria Pirzadeh, Vassilis~J. Tsotras, Rares Vernica, Jian Wen, and Till Westmann.
\newblock Asterixdb: {A} scalable, open source {BDMS}.
\newblock {\em Proc. {VLDB} Endow.}, 7(14):1905--1916, 2014.

\bibitem{DBLP:journals/pvldb/AlsubaieeBBHKCDL14}
Sattam Alsubaiee, Alexander Behm, Vinayak~R. Borkar, Zachary Heilbron, Young{-}Seok Kim, Michael~J. Carey, Markus Dreseler, and Chen Li.
\newblock Storage management in asterixdb.
\newblock {\em Proc. {VLDB} Endow.}, 7(10):841--852, 2014.

\bibitem{AsterixDBManual}
{Apache AsterixDB}.
\newblock {\em AsterixDB SQL++ User Manual}, 2023.

\bibitem{DBLP:journals/tods/AstrahanBCEGGKLMMPTWW76}
Morton~M. Astrahan, Mike~W. Blasgen, Donald~D. Chamberlin, Kapali~P. Eswaran, Jim Gray, Patricia~P. Griffiths, W.~Frank~King III, Raymond~A. Lorie, Paul~R. McJones, James~W. Mehl, Gianfranco~R. Putzolu, Irving~L. Traiger, Bradford~W. Wade, and Vera Watson.
\newblock System {R:} relational approach to database management.
\newblock {\em {ACM} Trans. Database Syst.}, 1(2):97--137, 1976.

\bibitem{DBLP:journals/pvldb/BalkesenATO13}
Cagri Balkesen, Gustavo Alonso, Jens Teubner, and M.~Tamer {\"{O}}zsu.
\newblock Multi-core, main-memory joins: Sort vs. hash revisited.
\newblock {\em Proc. {VLDB} Endow.}, 7(1):85--96, 2013.

\bibitem{DBLP:journals/tods/BittonD83}
Dina Bitton and David~J. DeWitt.
\newblock Duplicate record elimination in large data files.
\newblock {\em {ACM} Trans. Database Syst.}, 8(2):255--265, 1983.

\bibitem{DBLP:journals/tods/DoGN22}
Thanh Do, Goetz Graefe, and Jeffrey~F. Naughton.
\newblock Efficient sorting, duplicate removal, grouping, and aggregation.
\newblock {\em {ACM} Trans. Database Syst.}, 47(4):16:1--16:35, 2022.

\bibitem{DBLP:journals/csur/Graefe93}
Goetz Graefe.
\newblock Query evaluation techniques for large databases.
\newblock {\em {ACM} Comput. Surv.}, 25(2):73--170, 1993.

\bibitem{DBLP:conf/icde/Graefe94}
Goetz Graefe.
\newblock Sort-merge-join: An idea whose time has(h) passed?
\newblock In {\em Proceedings of the Tenth International Conference on Data Engineering, February 14-18, 1994, Houston, Texas, {USA}}, pages 406--417. {IEEE} Computer Society, 1994.

\bibitem{DBLP:journals/csur/Graefe06}
Goetz Graefe.
\newblock Implementing sorting in database systems.
\newblock {\em {ACM} Comput. Surv.}, 38(3):10, 2006.

\bibitem{DBLP:journals/ife/Graefe12}
Goetz Graefe.
\newblock New algorithms for join and grouping operations.
\newblock {\em Comput. Sci. Res. Dev.}, 27(1):3--27, 2012.

\bibitem{DBLP:journals/tkde/GraefeLS94}
Goetz Graefe, Ann Linville, and Leonard~D. Shapiro.
\newblock Sort versus hash revisited.
\newblock {\em {IEEE} Trans. Knowl. Data Eng.}, 6(6):934--944, 1994.

\bibitem{DBLP:journals/vldb/HaasCLS97}
Laura~M. Haas, Michael~J. Carey, Miron Livny, and Amit Shukla.
\newblock Seeking the truth about ad hoc join costs.
\newblock {\em {VLDB} J.}, 6(3):241--256, 1997.

\bibitem{DBLP:conf/vldb/Harder77}
Theo H{\"{a}}rder.
\newblock A scan-driven sort facility for a relational database system.
\newblock In {\em Proceedings of the Third International Conference on Very Large Data Bases, October 6-8, 1977, Tokyo, Japan}, pages 236--244. {IEEE} Computer Society, 1977.

\bibitem{DBLP:conf/sigmod/Jahangiri21}
Shiva Jahangiri.
\newblock Wisconsin benchmark data generator: To {JSON} and beyond.
\newblock In Guoliang Li, Zhanhuai Li, Stratos Idreos, and Divesh Srivastava, editors, {\em {SIGMOD} '21: International Conference on Management of Data, Virtual Event, China, June 20-25, 2021}, pages 2887--2889. {ACM}, 2021.

\bibitem{DBLP:phd/basesearch/Kim18a}
Taewoo Kim.
\newblock {\em Enhancing Apache AsterixDB for Efficient Big Data Search and Analytics}.
\newblock PhD thesis, University of California, Irvine, {USA}, 2018.

\bibitem{DBLP:journals/spe/KimBBBBCHJJLLMP20}
Taewoo Kim, Alexander Behm, Michael Blow, Vinayak~R. Borkar, Yingyi Bu, Michael~J. Carey, Murtadha~Al Hubail, Shiva Jahangiri, Jianfeng Jia, Chen Li, Chen Luo, Ian Maxon, and Pouria Pirzadeh.
\newblock Robust and efficient memory management in apache asterixdb.
\newblock {\em Softw. Pract. Exp.}, 50(7):1114--1151, 2020.

\bibitem{DBLP:conf/vldb/KitsuregawaNT89}
Masaru Kitsuregawa, Masaya Nakayama, and Mikio Takagi.
\newblock The effect of bucket size tuning in the dynamic hybrid {GRACE} hash join method.
\newblock In Peter M.~G. Apers and Gio Wiederhold, editors, {\em Proceedings of the Fifteenth International Conference on Very Large Data Bases, August 22-25, 1989, Amsterdam, The Netherlands}, pages 257--266. Morgan Kaufmann, 1989.

\bibitem{DBLP:conf/sigmod/0002SLLF15}
Ingo M{\"{u}}ller, Peter Sanders, Arnaud Lacurie, Wolfgang Lehner, and Franz F{\"{a}}rber.
\newblock Cache-efficient aggregation: Hashing is sorting.
\newblock In Timos~K. Sellis, Susan~B. Davidson, and Zachary~G. Ives, editors, {\em Proceedings of the 2015 {ACM} {SIGMOD} International Conference on Management of Data, Melbourne, Victoria, Australia, May 31 - June 4, 2015}, pages 1123--1136. {ACM}, 2015.

\bibitem{DBLP:conf/vldb/NakayamaK88}
Masaya Nakayama, Masaru Kitsuregawa, and Mikio Takagi.
\newblock Hash-partitioned join method using dynamic destaging strategy.
\newblock In Fran{\c{c}}ois Bancilhon and David~J. DeWitt, editors, {\em Fourteenth International Conference on Very Large Data Bases, August 29 - September 1, 1988, Los Angeles, California, USA, Proceedings}, pages 468--478. Morgan Kaufmann, 1988.

\bibitem{DBLP:conf/bigdataconf/PirzadehCW15}
Pouria Pirzadeh, Michael~J. Carey, and Till Westmann.
\newblock Bigfun: {A} performance study of big data management system functionality.
\newblock In {\em 2015 {IEEE} International Conference on Big Data {(IEEE} BigData 2015), Santa Clara, CA, USA, October 29 - November 1, 2015}, pages 507--514. {IEEE} Computer Society, 2015.

\bibitem{DBLP:journals/tods/Shapiro86}
Leonard~D. Shapiro.
\newblock Join processing in database systems with large main memories.
\newblock {\em {ACM} Trans. Database Syst.}, 11(3):239--264, 1986.

\end{thebibliography}
\end{document}